# Stopping Resistance Drift in Phase Change Memory Cells and Analysis of Charge Transport in Stable Amorphous Ge$_2$Sb$_2$Te$_5$

Md Tashfiq Bin Kashem*, Raihan Sayeed Khan*, ABM Hasan Talukder, Faruk Dirisaglik and Ali Gokirmak

Department of Electrical and Computer Engineering, University of Connecticut, Storrs, CT 06269, USA. *Equal contribution.
Email: ali.gokirmak@uconn.edu

*Abstract*—We stabilize resistance of melt-quenched amorphous Ge$_2$Sb$_2$Te$_5$ (a-GST) phase change memory (PCM) line cells by substantially accelerating resistance drift and bringing it to a stop within a few minutes with application of high electric field stresses. The acceleration of drift is clearly observable at electric fields > 26 MV/m at all temperatures (85 K – 300 K) and is independent of the current forced through the device, which is a strong function of temperature. The low-field (< 21 MV/m) I-V characteristics of the stabilized cells measured in 85 K – 300 K range fit well to a 2D thermally-activated hopping transport model, yielding hopping distances in the direction of the field and activation energies ranging from 2 nm and 0.2 eV at 85 K to 6 nm and 0.4 eV at 300 K. Hopping transport appears to be better aligned with the field direction at higher temperatures. The high-field current response to voltage is significantly stronger and displays a distinctly different characteristic: the differential resistances at different temperatures extrapolate to a single point (8.9×10$^{-8}$ ohm.cm), comparable to the resistivity of copper at 60 K, at 65.6 ± 0.4 MV/m. The physical mechanisms that give rise to the substantial increase in current in the high-field regime also accelerate resistance drift. We constructed field and temperature dependent conduction models based on the experimental results and integrated it with our electro-thermal finite element device simulation framework to analyze reset, set and read operations of PCM devices.

## I. Introduction

Spontaneous increase of resistance with time in amorphous phase of phase change materials, known as resistance drift, can cause erroneous inter-mixing of intermediate states in multi-level cells and thus act as a bottleneck for denser storage in PCM [1], [2]. A significant effort has been devoted so far to identify the cause of drift and minimize it [3]–[8], however a comprehensive solution has not been produced yet. There is also a number of different transport mechanisms proposed for amorphous phase change materials [9]–[13]. In this work, (i) we experimentally demonstrate substantial acceleration of resistance drift and stabilization of device resistance in melt-quenched a-GST line-cells with application of high-field stresses (> 26 MV/m) in 85 K to 200 K range, (ii) characterize field and temperature dependent current conduction in stabilized devices, (iii) construct a 2D temperature dependent hopping transport model for low-field regime and extract the hopping distances, hopping angles and activation energies associated with percolation transport and (iv) construct an empirical model for electronic conductivity in the high-field regime. We integrate the resulting electric field and temperature dependent electrical

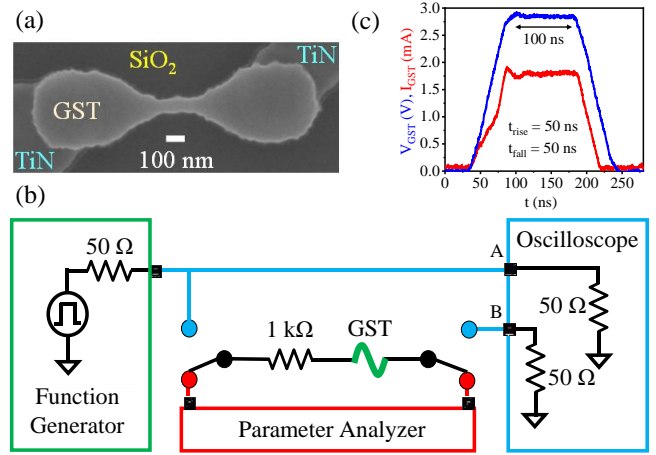

**Fig. 1** (a) SEM image of a GST line cell with TiN bottom contacts. (b) Electrical measurement setup. (c) Voltage across the GST wire (—) and measured current through the wire (—) during reset pulse.

conductivity for stable a-GST with our finite element simulation framework to model reset, set and read operations of PCM devices.

## II. Device fabrication and Characterization

GST line cells used for this study are fabricated by co-sputtering from elemental targets on thermally grown SiO$_2$ atop Si substrates with tungsten back contacts, patterned using photo-lithography and reactive ion etching, and capped by Si$_3$N$_4$, as described in [14] (Fig. 1a). The dimensions of the cells are: length (*l*) × width (*w*) × thickness (*th*) = ~600-700 nm × ~70-150 nm × ~50±5 nm. The cells are first crystallized to the hexagonal close packed (hcp) phase by annealing at 675 K and then amorphized using a single 100 ns pulse with 50 ns rise and fall times (Fig. 1b,c) in 85 K to 300 K temperature range in a Janis ST-500-UHT cryogenic probe station under vacuum (~0.01 mTorr). Pulse width and rise/fall times are chosen to minimize reflections and parasitic contributions in the measurement setup while ensuring amorphization of the cells without substantial distortion of the waveforms. After the amorphizing reset pulse, five DC I-V sweeps are performed at each temperature using an Agilent 4156C parameter analyzer (0 V to 25 V and back to 0 V in 0.1 V steps) with current compliance set to 50 nA. The width of the devices in the analysis and the construction of the models is ~152 nm.

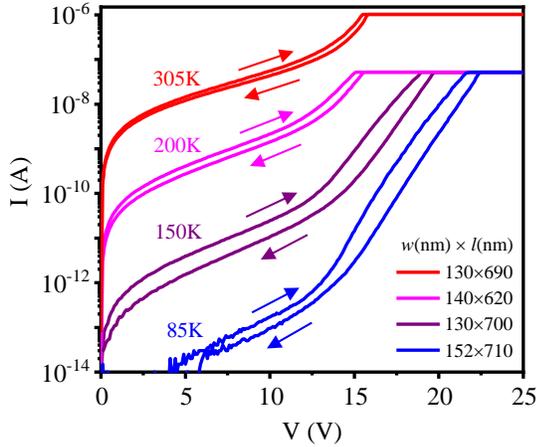

**Fig. 2** I-V sweeps immediately after amorphization for four different devices at indicated temperatures. Device dimensions are indicated in nm as $w \times l$. Current compliance was set to 1 µA for 305K and 50 nA for the rest.

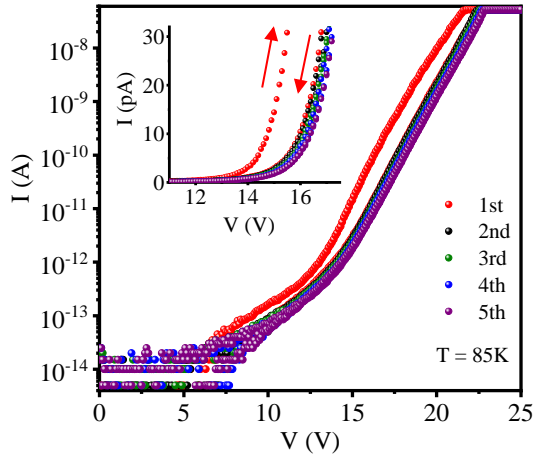

**Fig. 3** All five I-V sweeps after amorphization at 85K. Inset shows a zoomed version of the plot in 11V to 17.5V voltage range, the up and down arrow indicate the upswing and downswing currents of the first I-V sweep.

### III. STOPPAGE OF RESISTANCE DRIFT

The first I-V sweeps after reset show a clear hysteresis behavior with larger hysteresis windows at lower temperatures (Fig. 2). The subsequent sweeps for T < 200K display significantly smaller hysteresis windows and devices stabilize within 3 sweeps (Fig. 3). We observe a stronger response to electric field above ~21 MV/m, transitioning from a low-field response to a high-field response, and substantial acceleration in resistance drift at ~26 MV/m. Devices reach their final resistance value within minutes with high electric-field stress, which normally takes place in months without stress (Fig. 4). As the current compliance is small (50 nA), self-heating and thermally induced structural relaxation are not expected in the devices as a whole. However, filamentary conduction is expected and higher temperatures may be reached along these filaments. The stress induces changes on the percolation paths, device resistance increases and fluctuations in current decrease (Fig. 5). These changes may be due to relaxation of the charges left in the traps within a-GST as devices quench [4] or annihilation of unstable trap sites that assist percolation transport. The first being only due to charge relaxation and the second being due to movement of unstable atoms in the structure. Both of these processes would lead to substantial distortion of the potential profiles and change the trapped-charge emission rates.

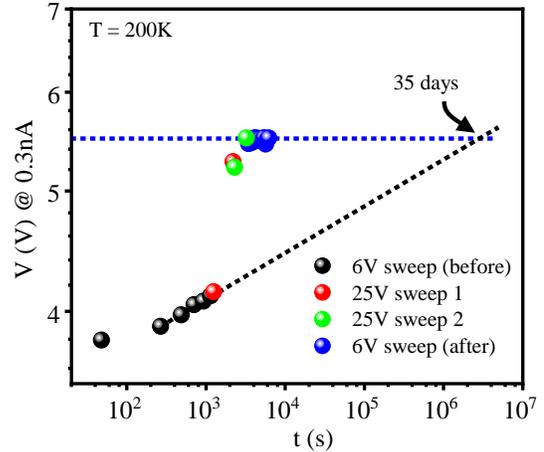

**Fig. 4** Bi-logarithmic plot of voltage required for current to reach 0.3 nA versus time at 200K. Required voltage increases following a power law behavior (●) before the application of 25V sweep (●), after which the voltage increases substantially (acceleration of drift) and becomes stable (●) afterwards (stoppage of drift).

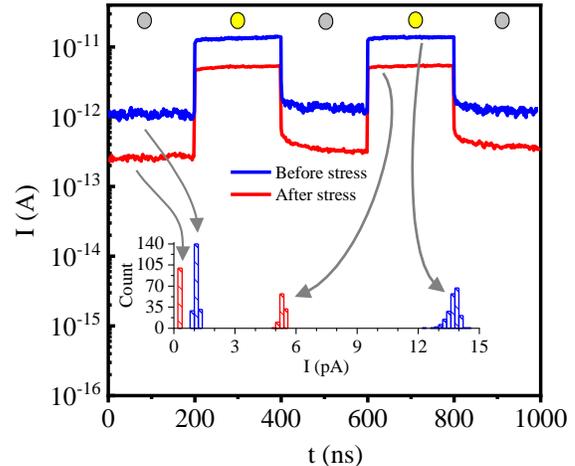

**Fig. 5** Current versus time plot showing the effect of electrical stress and light at 200K. The yellow and grey circles indicate the LED being on and off during that time period. Inset shows the histograms before and after the application of high voltage stress from 0 s to 200 ns time period under dark condition and in 600 ns to 800 ns time interval under light. The device dimensions are: $l \sim 690$ nm and $w \sim 146$ nm.

### IV. MODELING CHARGE TRANSPORT IN STABLE AMORPHOUS GST

#### A. Low field transport

We observe two distinct exponential responses in the I-V characteristics of the stable cells in the low-field and high-field regimes (Fig. 7). Low-field response can be modeled as thermionic emission over a barrier with an activation energy ($E_a$), where the symmetry between the forward transmission and reverse transmission is broken by the external field; the barrier seen by the trapped carriers ($E_{peak}$) is modulated by the external field (Fig. 6). The changes in the potential energy of the barrier peak ($\Delta E_{peak}$), the source-trap ($\Delta E_{source}$) and the destination-trap ($\Delta E_{destination}$) as a function of the external bias (V) depend on the profile of the barrier (Fig. 6b,c), the amorphized length (L) and trap-to-trap distance ($d_{trap}$), and can be expressed using geometric relationships assuming a uniform field profile throughout the device in 1D:

$$I_{Forward} = I_f\, e^{-\{E_a - (\Delta E_{peak} - \Delta E_{source})\}/k_B T} \quad (1)$$

$$I_{Reverse} = I_r\, e^{-\{E_a - (\Delta E_{peak} - \Delta E_{destination})\}/k_B T} \quad (2)$$

$$\Delta E_{destination} - \Delta E_{source} = qV \frac{d_{trap}}{L} \quad (3)$$

$$\Delta E_{peak} - \Delta E_{source} = qV \frac{d_{peak}}{L} = qV \frac{b\, d_{trap}}{L} \quad (4)$$

$$\Delta E_{destination} - \Delta E_{peak} = qV \frac{d_{trap} - d_{peak}}{L} = qV \frac{(1-b)d_{trap}}{L} \quad (5)$$

where $k_B$ is the Boltzmann constant, q is the elementary charge, $d_{peak}$ is the location of the barrier peak measured from the source-trap and $b = d_{peak}/d_{trap}$ is the relative position of the peak: $b = 0$ represents a barrier peak at the source-trap (Fig. 6b) and $b = 1$ represents a barrier peak at the destination-trap. The net current can be expressed as a combination of forward and reverse thermionic emission over a biased barrier:

$$I_{Low\,field} = I_f\, e^{\frac{-E_a + qV \frac{b\, d_{trap}}{L}}{k_B T}} - I_r\, e^{\frac{-E_a - qV \frac{(1-b)d_{trap}}{L}}{k_B T}} \quad (6)$$

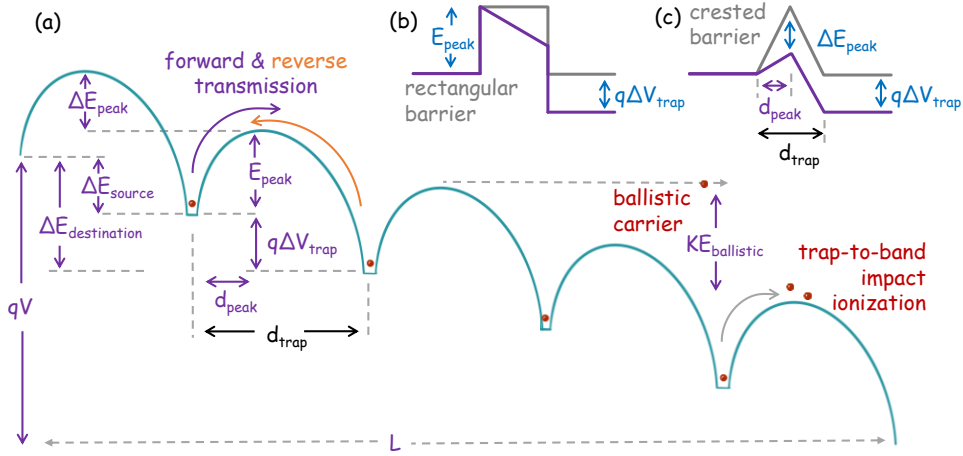

**Fig. 6** A schematic illustration of conduction-band edge and n-type charge transport in a material with traps. The net carrier flow can be modeled using the forward and reverse transmission probabilities (a). Thermionic emission probabilities will depend on the shape of the energy barrier between the trap sites. The barrier peak appears at the source side (left) for a rectangular barrier (b), whereas the peak is in the center for a simple crested (or triangular) barrier (c). The peak height of the energy barrier ($E_{peak}$) does not change for a rectangular barrier, while it is modulated ($\Delta E_{peak}$) by the applied bias (V) for a crested barrier. The modulation in the barrier height depends on the ratio of the peak location ($d_{peak}$) and trap separation ($d_{trap}$). Carriers can be accelerated to sufficiently high kinetic energies ($KE_{ballistic}$) to impact ionize trapped charges, or give rise to band-to-band excitation, depending on V.

Defining $\omega = q d_{trap}/k_B T L$ and assuming that the carrier concentration available for forward and reverse transmission are equal at the rate limiting trap-sites ($I_f = I_r = I_m$), and b is insensitive to the applied bias, the low-field current expression can be simplified as:

$$I_{Low\,field} = I_m\, e^{\frac{-E_a}{k_B T}} \left\{ e^{b\omega V} - e^{-(1-b)\omega V} \right\} \quad (7)$$

$$I_{Low\,field}(V,T) = I_0(T)\left[ e^{b(T)\omega V} - e^{-\{1-b(T)\}\omega V} \right] \quad (8)$$

Fitting the experimental data to (8) in the whole temperature range, in the 0-10V voltage range (Fig. 7), we obtain b, the pre-factor ($I_0$), $d_{trap}$, and $d_{peak}$ as a function of temperature (Fig. 8a-c). With the extrapolations made for the higher temperatures, we can predict the conductivity (σ) and carrier activation energies ($E_a$) (Fig. 8d,e) at higher temperatures. We observe b(300 K) = $d_{peak}/d_{trap}$ ≈ 0.54 (i.e. barrier peak is close to the center point between the traps). This is inline with the report by Ielmini et al [15] for their 1D hopping transport model developed to analyze the results at and above room temperature. In their model, the barrier is assumed to be in the middle and the barrier height is modulated by half of the voltage difference between the trap sites (Fig. 6c). We also note that our $d_{trap}$ (300 K) ≈ 6nm (Fig. 8c) is close to the room temperature inter-trap distance reported in another work of Ielmini et al [8].

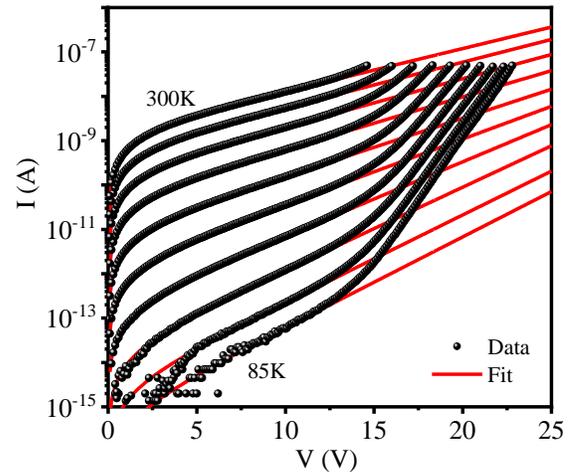

**Fig. 7** Average of last four I-V measurements at 85K and 100 K to 300K with 25 K steps, after amorphization at 85 K (●) and fitted I-V behavior with the model described by (8) in 0V to 10V range and extrapolated to the full range (—).

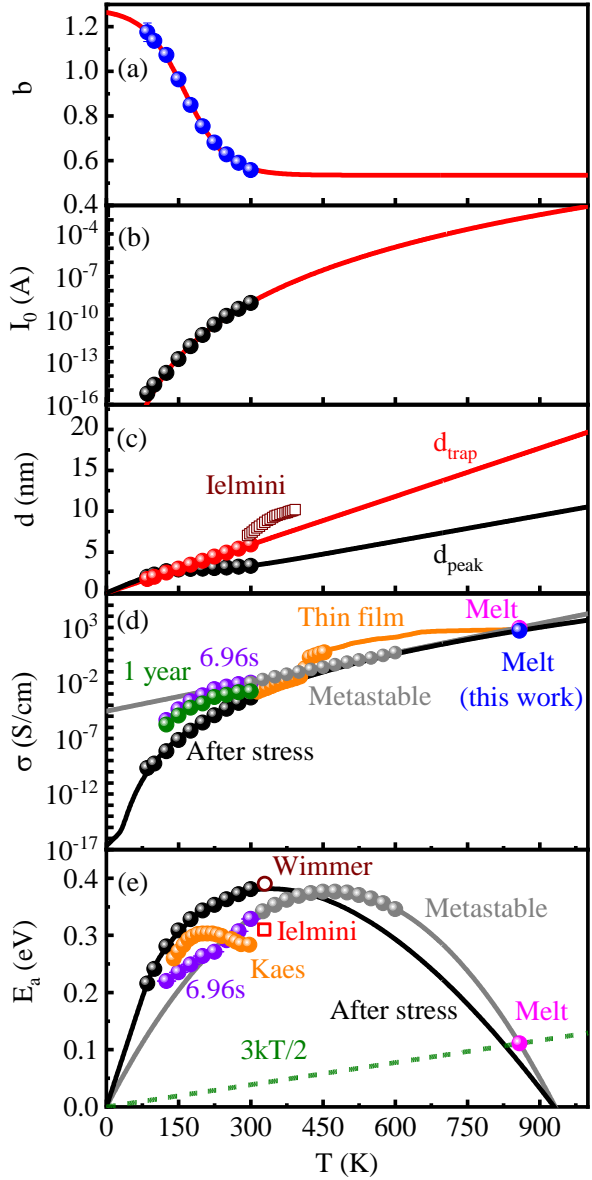

**Fig. 8** Temperature dependence of low field transport parameters. Fit parameters: (a) b(T) (●), and (b) $I_0$ (T) (●) obtained from the fits to I-V data of the stable a-GST in the low field region (0V-10V). The red line (—) in (a) shows saturating exponential fit to the symbols and power law fit to the data in (b). (c) Calculated energy barrier peak location in reference to the source-trap ($d_{peak}$) and calculated trap-to-trap distance ($d_{trap}$) as a function of temperature, calculated $d_{trap}$ values by Ielmini et al. [8]. (d) Low-field conductivity (I/V) at 3V after stress (●) calculated from the experimental I-V data, modelled using b(T) and $I_0$(T) (—), predicted metastable (●) and 1 year drifted (●) conductivity based on slow measurements before stress, conductivity measured from: metastable wire [14] (●) and corresponding fit [16] (—), thin-film continuous R-T sweep (—), Seebeck coefficient-T measurements [17] (●) and melt [18] (●), melt (●) calculated from the amorphization pulse I-V shown in Fig. 1c. (e) Carrier activation energy after high-field stress calculated using (10) (● and —), metastable values [16] (● and —), values from the literature [7], [10], [15], and 3kT/2 line indicating the average kinetic energy of the particle (--).

We calculated drifted low-field conductivity at 3V using (8) and find it to follow the drifted thin-film conductivity of as deposited a-GST [17] for T < $T_g$ and melt-quenched metastable a-GST conductivity [14] for T > $T_g$ (Fig. 8d) where $T_g$ ≈ 407K is the glass transition temperature. Charge carrier activation energy can be determined using:

$$I_m e^{-E_a(T)/k_B T} = I_0(T) \quad (9)$$

$$E_a(T) = k_B T \ln[I_m / I_0(T)] \quad (10)$$

Here, we need a reference activation energy to calculate $I_m$, which may be $E_a$ ($T_{metal}$)= 0 eV, where $T_{metal}$ ~ 930K is the semiconductor to metal transition temperature we had previously calculated for liquid GST [16]. $E_a$ for stressed a-GST we calculate is comparable to the previously reported values around room temperature [7] (Fig. 8e), larger than the metastable $E_a$ [16] below ~$T_g$.

Temperature dependence of b ($d_{peak}/d_{trap}$, sensitivity of the barrier height to applied bias) increases with reduced T (Fig. 8a), indicating that the barrier peak appears closer to the destination trap at lower temperatures. However, b exceeds 1 for T ~125 K, corresponding to a barrier maximum position beyond the destination-trap location, which is not physical in 1D. However, in 2D and 3D percolation transport, hopping path of the carriers do not align with the externally applied field. The percolation path at the rate-limiting sites and the external field may have a large angle between them ($\theta_{trap}$). Assuming that the carriers go around 2D potential peaks on their percolation paths in a circular fashion, $d_{peak}$ along the direction of external field can be longer than $d_{trap}$ along the direction of the external field (Fig. 9). Assuming electric field, $\vec{E}$ is along the x-direction, and $\theta_{trap}$ is the angle between $\vec{E}$ and $\vec{d_{trap}}$, b can be expressed as:

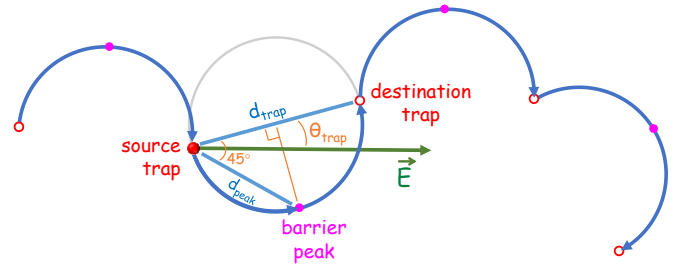

**Fig. 9** Schematic diagram of 2D low field transport under externally applied electric field, $\vec{E}$ (—) assuming circular hopping path for trapped carrier (●) from source trap-site to the destination one where the barrier peaks are shown by (●).

$$b = \frac{\overrightarrow{d_{peak}} \cdot x}{\overrightarrow{d_{trap}} \cdot x} \quad (11)$$

$$b = \frac{(d_{trap}/\sqrt{2})\cos(\theta_{trap} - 45°)}{d_{trap}\cos(\theta_{trap})} \quad (12)$$

Using (12), we obtain a relationship between $\theta_{trap}$ and b:

$$b = \frac{1}{\sqrt{2}} \frac{\cos(\theta_{trap})\cos(45°) + \sin(\theta_{trap})\sin(45°)}{\cos(\theta_{trap})} \quad (13)$$

$$b = \frac{1}{\sqrt{2}} \frac{1}{\sqrt{2}} \frac{\cos(\theta_{trap}) + \sin(\theta_{trap})}{\cos(\theta_{trap})} \quad (14)$$

$$b = \frac{1}{2}\left[1 + \tan(\theta_{trap})\right] \quad (15)$$

$$\theta_{trap} = \tan^{-1}(2b - 1) \quad (16)$$

Using the b(T) obtained from the experimental data (Fig. 8a), we calculate $\theta_{trap}$ to decrease as a function of temperature, asymptotically approaching 4° around room temperature (Fig. 10a) and $\theta_{trap}$ exceeds 45° for T < 125 K for this particular device. Based on the extrapolations we made, $\theta_{trap}$ is expected to remain < 60° for this device. The 2D model predicts a slightly larger $d_{trap}$ at low temperatures compared to the 1D model (Fig. 10b). Due to the circular path assumption, $d_{peak} = d_{trap}/\sqrt{2}$ for all temperatures.

### B. High field transport

The significant change that we observe above ~21 MV/m may be attributed to carriers starting to follow shorter percolation paths with the barriers lowered by the external bias, the activation of deeper traps or impact ionization of trapped charges. We observed that the differential resistances, $R_d = dV/dI$, at high fields follow a simple exponential for all temperatures. Fitting the natural logarithm of differential

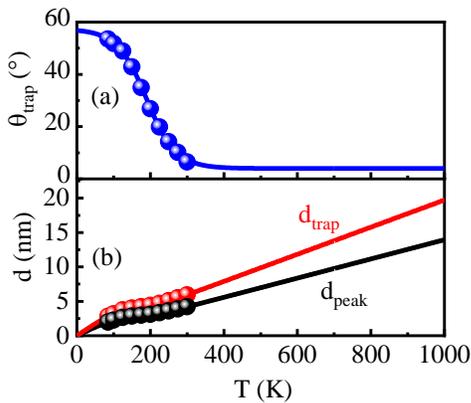

**Fig. 10** (a) $\theta_{trap}$ and (b) $d_{trap}$ and $d_{peak}$ as a function of temperature obtained from the circular transport model analysis.

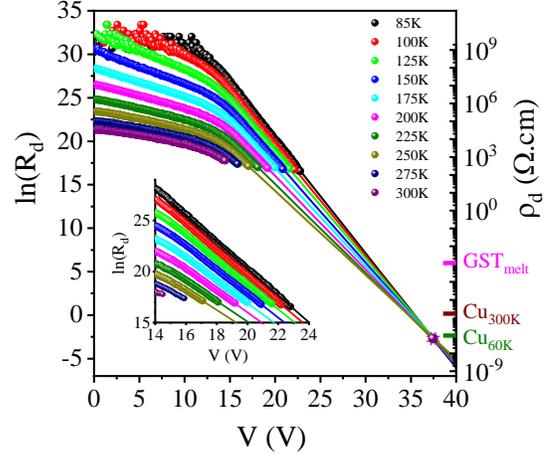

**Fig. 11** Natural logarithm of the differential resistance, $\ln(R_d)$ and differential resistivity, $\rho_d$ as a function of voltage. The lines demonstrate the fits made to the $\ln(R_d)$ data in the high field regime at different temperatures. 275K and 300K cases do not have sufficient high field data to do the fits. The inset shows zoomed version of the data and fits in the fitting range. All the fits converge approximately at 65.6 MV/m (●). $\rho_d$ of GST at the melting temperature, $T_{melt}$ and Cu at 60K and 300K are taken from refs. [17] and [19].

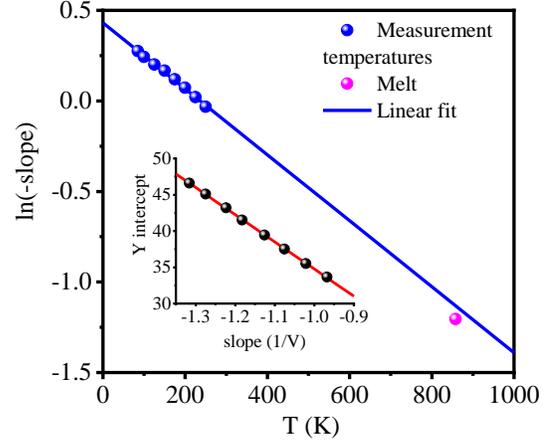

**Fig. 12** Natural logarithm of the slopes of the linear fits of Fig. 11, which are fitted to a line and the inset shows the linear relationship between the intercepts and slopes of the linear fits of Fig. 11.

resistances to a simple linear model, we obtain the temperature dependence of the slope and the intercept (Fig. 11):

$$\ln(R_d) = m(T)V + c(T) \quad (17)$$

The linear fits made to the data converge to a single point for all temperatures at 65.6 ± 0.4 MV/m, indicating that transport becomes temperature independent at that point. Differential resistivity at that point is comparable to the resistivity of Copper (Cu) at 60 K [19]. It should also be noted that we do not observe such a convergence in the low-field regime.

The slope m(T) appears to follow a simple exponential function of T (Fig. 12) and y-intercept, c(T) varies linearly with m(T) (Fig. 12 inset):

$$m(T) = Pe^{ST}, \quad c(T) = m(T)A + B \quad (18)$$

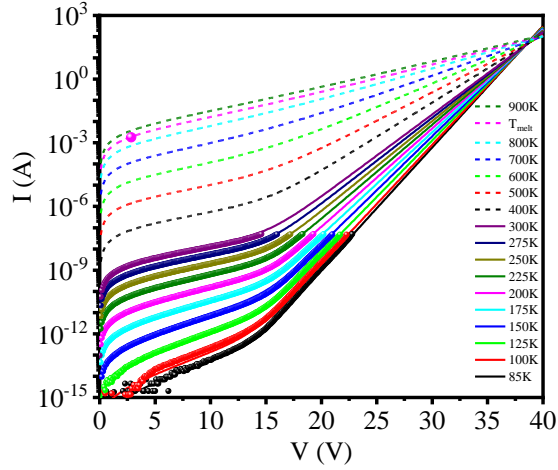

**Fig. 13** Experimental I-V at different temperatures (spheres), modelled I-V at the measurement temperatures (solid lines), and beyond the measurement temperatures (dashed lines).

where P is the pre-factor, S is the exponent, A and B are slope and y-intercept of the respective fits (Fig. 12). Finally, current in the high field-regime can be expressed as:

$$I_{High\ field}(V,T) = \int_0^V \frac{1}{R_d(V,T)}dV = \frac{e^{-c(T)}}{-m(T)}\left[e^{-m(T)V} - 1\right] \quad (19)$$

$$I_{High\ field}(V,T) = \frac{1}{P}e^{-B-ST}\left[e^{-APe^{ST}} - e^{-P(V+A)e^{ST}}\right] \quad (20)$$

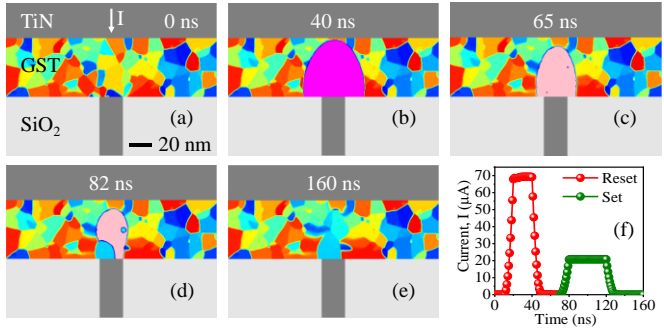

**Fig. 14** Finite element electro-thermal modeling of reset and set operation of a typical PCM mushroom cell utilizing the extracted electric field and temperature dependent conductivity of the amorphous GST, $\sigma_{a\text{-}GST}$ from the I-V model of Fig. 13. Time evolution of the phase distribution maps are shown: before reset (crystalline phase) (a), at the peak of melting (b), after quenching of the molten phase resulting in amorphous phase (c), during set (d) and at the end of set (back to crystalline phase again with different grain orientations than the initial crystalline phase) (e). TiN is used as the top and bottom electrode (heater) while $SiO_2$ works as the insulator. An n-channel metal oxide semiconductor field effect transistor (MOSFET) is used as the access device (not shown in the figure). Current through the device, I is shown in (f) as a function of time. Simulations are based on the framework described in [23], [24] using COMSOL Multiphysics platform [25]. Material parameters dynamically update based on electric field, temperature and phase variations with time. In the model, $\sigma_{a\text{-}GST}$ affects the electric current density as well as the electronic component of the thermal conductivity of the bulk GST and thermal boundary resistance at GST-TiN and GST-$SiO_2$ interfaces.

Linear combination of the low-field (8) and high-field (20) I-V models match the data very well in a large temperature and voltage range (Fig. 13). The conductance we calculate using (8) and (20) is implemented in our finite element simulation framework [20]–[24], where we model phase change and solve current continuity and heat transport equations self-consistently to perform electro-thermal simulation of reset, set and read operations of PCM devices (Fig. 14).

## V. CONCLUSION

In summary, we apply high electric field stress in melt-quenched a-GST PCM line cells to accelerate and stop resistance drift. The low-field conduction can be modeled using a 2D percolation hopping transport model. We observe a larger deviation from a straight path between the two device contacts at lower temperatures. The high-field conduction shows distinct characteristics and convergence of differential resistance to a single point for all temperatures. At this field (65.6 ± 0.4 MV/m) differential resistance is temperature independent and may be considered the breakdown field for the material. The current versus voltage curves for different temperatures also converge at a slightly higher field. However, this convergence point is not as tight as the convergence point for differential resistances. The mechanisms giving rise to increased current in high-field regime also give rise to acceleration of resistance drift. Ability to accelerate and stop resistance drift can pave the way for reliable implementation of multi-bit per cell storage.

### ACKNOWLEDGEMENT

Author contributions: Kashem and Gokirmak performed the data analysis and construction of the models. Khan and Gokirmak performed the experimental studies and did the first observation of acceleration and stoppage of resistance drift. Talukder performed additional drift experiments and data analysis. Dirisaglik fabricated the devices at IBM T. J. Watson Research Center.

Authors are grateful for the contributions of IBM staff for device fabrication. This work is supported by US NSF ECCS grant # 1711626.